\documentclass[10pt, conference]{IEEEtran}
\usepackage{algorithmicx}
\usepackage[ruled,vlined,linesnumbered]{algorithm2e}
\usepackage{cite}
\usepackage{hhline}
\usepackage{amsmath,mathtools}
\usepackage{amsfonts,amssymb}
\usepackage{mathrsfs}
\usepackage{caption}
\usepackage{multirow}
\usepackage{graphicx} 
\usepackage{multirow}
\usepackage{geometry}
\usepackage{enumitem,color}
\usepackage{algpseudocode}
\begin{document}
%
\title{\huge \emph{CAVE}: Crowdsourcing Passing-By Vehicles for Reliable In-Vehicle Edge Computing \vspace{-0.1in}}

\author{\IEEEauthorblockN{Jiahe Cao, Qiang Liu \vspace{-0.16in}}\\
\IEEEauthorblockA{
University of Nebraska-Lincoln\\
qiang.liu@unl.edu}\vspace{-0.42in}
\and
\IEEEauthorblockN{Dawei Chen, Kyungtae Han \vspace{-0.16in}}\\
\IEEEauthorblockA{
Toyota InfoTech Labs\\
\{dawei.chen1, kt.han\}@toyota.com}\vspace{-0.42in}
}

\maketitle

\begin{abstract}
In-vehicle edge computing is a much anticipated paradigm to serve ever-increasing computation demands originated from the ego vehicle, such as passenger entertainments.
In this paper, we explore the unique idea of crowdsourcing passing-by vehicles to augment computing of the ego vehicle.
The challenges lie in the high dynamics of passing-by vehicles, time-correlated task computation, and the stringent requirement of computing reliability for individual user tasks.
To this end, we formulate an optimization problem to minimize the end-to-end latency by optimizing the task assignment and resource allocation of user tasks.
To address the complex problem, we propose a new algorithm (named \emph{CAVE}) with multiple key designs.
First, we reformulate the original problem into two subproblems while incorporating not only incoming but also in-progress tasks.
Second, we solve the task assignment subproblem with reliability constraints by using particle swarm optimization with the adaptive barrier function. 
Third, we solve the resource allocation subproblem by deriving the optimal allocation with Karush–Kuhn–Tucker (KKT) condition.
We build an end-to-end network and compute simulator and conduct extensive simulation to evaluate the performance of the proposed algorithm.
Simulation results show that, our \emph{CAVE} algorithm reduces more than 15\% end-to-end latency than state-of-the-art solutions, without degrading the reliability performance.
\end{abstract}


\section{Introduction}
\label{sec:introduction}

Vehicle computing~\cite{lu2023vehicle} is much anticipated to be the next edge computing paradigm to enable and catalyze a wide range of applications on the drive, such as augmented and virtual reality (AR/VR) and large language models (LLMs) from passengers~\cite{mao2023gpt}.
With the electrification and intelligentization of vehicles, e.g., autonomous driving (AD)~\cite{yurtsever2020survey} and software-defined vehicles (SDV)~\cite{jiacheng2016software}, modern vehicles are with much higher computation, networking, and storage capabilities, for example, NVIDIA DRIVE AGX delivers 200 TOPS.
Apart from the needed capability to perform vehicular-related computation (if enabled), e.g., AD and advanced driver-assistance system (ADAS), there remains non-trivial but varying computation and networking resources~\cite{liu2022prophet}, which can be utilized for accelerating generic computation tasks, such as AR/VR rendering and DNN inference.
This creates a unique possibility to leverage the sparse computation capability of vehicles as the edge computing servers, towards pervasive vehicle edge computing.

Vehicle edge computing has been widely investigated from various perspectives~\cite{liu2019edge, raza2019survey, ma2021parking}, including computation offloading, resource allocation, content caching, security, and privacy.
As high-mobility vehicles have extremely high dynamics (especially wireless connectivity on the drive), one of the key considerations is the reliability of vehicle computation~\cite{hou2020reliable, liu2023accelerating}, in terms of completing user tasks. 
For example, a vehicle may be disconnected from wireless networks (e.g., cellular vehicle-to-everything (C-V2X)), or experience a significantly low wireless data rate, which can fail its in-progress tasks.
Recent works~\cite{hou2020reliable, liu2023accelerating} investigated assigning user tasks to multiple vehicles for increasing computation redundancy and thus improving computing reliability.
In existing reliability works, all vehicles are connected via the common wireless network infrastructure (e.g., C-V2X Uu and public cellular), where vehicles receive user tasks from the network side, and send computing results back after task completion, via one-hop wireless transmission (either vehicle-to-infrastructure or infrastructure-to-vehicle).
However, these approaches can be problematic when user tasks originate from individual vehicles (e.g., passengers), which will involve two-hop wireless transmission (i.e., vehicle-to-infrastructure-to-vehicle and reverse path) and result in highly varying and delayed end-to-end performance.

In this paper, we focus on the unique idea of crowdsourcing passing-by vehicles to augment computing of the ego vehicle, where user tasks originate from the ego vehicle.
The rationale is that, there are more computation demands originating from vehicles, including but not limited to, passenger entertainment (e.g., AR/VR), personalized agents (e.g., LLMs), and enhanced driving (e.g., connected AD). 
In this scenario, the ego vehicle wirelessly connects with passing-by vehicles and sends user tasks to them for reliable computation, if not locally computed. 
Here, the unique challenges lie in the high dynamics of passing-by vehicles, time-correlated task computation, and the stringent requirement of computing reliability for individual user tasks.

To address these challenges, we formulate an optimization problem to optimize the task assignment and resource allocation of user tasks, where each task can be assigned to multiple passing-by vehicles to improve computing reliability.
Specifically, we aim to minimize the average end-to-end latency of tasks originating from the ego vehicle, while ensuring the requirement of computing reliability of individual tasks.
We propose a new algorithm (named \emph{CAVE}) to efficiently solve the optimization problem, with the following key designs. 
1) We reformulate the original problem into two subproblems at different locations, i.e., task assignment in the ego vehicle and resource allocation in individual passing-by vehicles.
In particular, we tackle the time-correlation issue by incorporating the existing in-progress tasks into both subproblems, where we do not modify their task assignments but re-optimize their resource allocation.
2) We convert the task assignment subproblem into unconstrained by using the adaptive barrier function, and solve the converted problem by using enhanced particle swarm optimization. 
3) We solve the resource allocation subproblem by deriving the optimal allocation with the Karush–Kuhn–Tucker (KKT) condition.
Note that, the algorithm does not require full controllability of the wireless network (which is more practical in real-world scenarios), and can be easily adapted to different wireless connectivities.
Extensive simulation results show that our proposed algorithm substantially outperforms existing solutions, in terms of convergence, reliability, and scalability.


\section{System Model}







\begin{figure}[!t]
	\centering
	\includegraphics[width=3in]{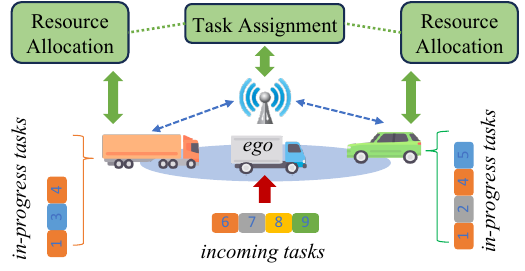}
	\caption{Overview of in-vehicle edge computing.}
	\label{fig:overview}
\end{figure}

We consider a generic vehicle edge computing network shown in Fig.~\ref{fig:overview}, including an ego vehicle and multiple passing-by vehicles.
The ego vehicle serves as an edge computing platform to accelerate the task computation from its passengers (e.g., AR/VR headset and mobile gaming).
We consider the passengers' devices to be connected with the ego vehicle, either wireless (WiFi) or wired (plug-in), with extremely high and consistent data rate, and thus omit the modeling of such transmission latency.
When the ego vehicle drives on the road, it crowdsources passing-by vehicles via direct wireless connectivity (e.g., C-V2X PC5 interface) to accelerate its task computation.
Due to the high dynamics of passing-by vehicles, we aim to assure the computation reliability of tasks by assigning individual tasks to one or more vehicles. 
As tasks are completed in passing-by vehicles, they will be sent back to the ego vehicle, and then forwarded to the devices of passengers.
We denote $\mathcal{I}$ as the computation tasks and $\mathcal{J}$ as the participating vehicles.
We define $\alpha_{i,j}$ as the binary indicator of task assignment, where $\alpha_{i,j}=1$ means the $i$th task is assigned to the $j$the vehicle.
Moreover, we define $g_{i,j}$ as the allocated computation resources to the $i$th task by the $j$th vehicle.
Denote $\mathcal{A} = \{ \alpha_{i,j}, \forall i, j\}$ and $\mathcal{G} = \{ g_{i,j}, \forall i, j\}$, which are the optimization variables.

\textbf{Reliability Model.}
From the perspective of the ego vehicle, other passing-by vehicles are highly dynamic throughout its driving period.
Assigning a task to only one passing-by vehicle may be unreliable, e.g., vehicles drive away from the coverage of the ego vehicle before they complete the assigned tasks.
To ensure the reliable computation of tasks, we allow each task to be assigned to one or more passing-by vehicles.
We consider, each passing-by vehicle has a reliability function~\cite{hou2020reliable} (denoted as $P_j(\cdot), \forall j$), which represents its probability of failing task computation (e.g., out of the coverage of the ego vehicle and other connectivity issues).
Note that, this reliability function is not constant but varies over time.
Then, we can calculate the reliability of computing a task in the $j$th vehicle as $P_j(L_{i,j})$, where $L_{i,j}$ is the round-trip latency of the task (see Eq.~\ref{eq:latency}).
For the sake of simplicity, we consider that the reliability function of all passing-by vehicles are known when solving the following optimization problem.
Hence, given the task assignment $\alpha_{i,j}, \forall j$, we express the unreliable probability of the $i$th task when it is computed by multiple vehicles as
\begin{align}
    U_i = \prod\nolimits_{j\in \mathcal{J}} \left(1-\alpha_{i,j}P_j(L_{i,j}) \right). \label{eq:unreliability}
\end{align}

\textbf{Latency Model.}
Each task experiences three stages, i.e., downlink ego-to-vehicles transmission, vehicle computation, and uplink vehicles-to-ego transmission, where both transmission stages may be omitted if the task is locally computed by the ego vehicle.
Hence, we model the round-trip latency~\cite{xue2023comap} of $i$th task in the $j$th vehicle as 
\begin{align}
    L_{i,j} = {D_i}/{R^d_{i,j}} + {C_i}/{g_{i,j}} + {E_i}/{R^u_{i,j}}, \label{eq:latency}
\end{align}
where $R^u_{i,j}$ and $R^d_{i,j}$ are the uplink (from other vehicles to the ego vehicle) and downlink (from the ego vehicle to other vehicles) wireless data rate.
Here, we do not assume that the ego vehicle has full controllability over the wireless communication, which is independently managed by other parties, e.g., cellular network operators.
In other words, the downlink $R^d_{i,j}$ and uplink $R^u_{i,j}$ wireless data rate experienced by the $i$th task at the $j$th vehicle are generally unknown.
Moreover, $C_i$, $D_i$, and $E_i$ are the known computation complexity (e.g., GFLOP), downlink and uplink data size, respectively. 

\textbf{Problem.}
The objective is to minimize the aggregated latency of all the incoming tasks.
Therefore, we formulate the optimization problem $\mathbb{P}_0$ as follows:
\begin{align}
    \mathbb{P}_0: \;\;  \min _{\mathcal{A}, \mathcal{G}} \;\;\;\;& \sum\nolimits_{i \in \mathcal{I}} \sum\nolimits_{j \in \mathcal{J}} \alpha_{ij} L_{i,j} \\
    \text { s.t. } \;\;\;\;& \sum\nolimits_{i \in \mathcal{I}} \alpha_{ij} g_{i,j} \leq G_j^{\max}, \forall j, \label{eq:const_1} \\ 
                   \;\;\;\;& \prod\nolimits_{j \in \mathcal{I}} (1-\alpha_{i,j}P_j(L_{i,j}))  \leq H_i^{min}, \forall i, \label{eq:const_2} 
\end{align}
where $G_j^{\max}$ is the computation capacity of the $j$th vehicle and $H_i^{min}$ is the threshold of failure probability for the $i$th task.
The optimization variables are the task assignment $\mathcal{A} = \{\alpha_{i,j}, \forall i, j\}$ (binary) for each task and resource allocation $\mathcal{G} = \{g_{i,j}, \forall i, j\}$ (continuous) in each vehicle.
The first constraint in Eq.~\ref{eq:const_1} ensures that, the allocated computation resources to all tasks complies with the capacity $G_j^{\max}$ of individual vehicles.
The second constraint in Eq.~\ref{eq:const_2} ensures that, the failure probability of each task is below the given threshold $H_i^{min}$.

\textbf{Challenges.}
The technical challenges of addressing the above problem are multi-fold.
First, the optimization variables $\mathcal{A}$ and $\mathcal{G}$ are closely coupled in both the objective function and the second constraint.
Moreover, multiple parameters in the latency and reliability model, i.e., the experienced wireless data rate, are unknown, when solving the above optimization problem. 
Second, the vehicle edge computing system is tightly time-correlated, where the life-cycle of tasks spans non-negligible period, while incoming tasks could arrive at any time.
In other words, optimizing the incoming tasks could degrade the performance of existing in-progressing tasks, which leads to complicated far-reaching impacts over long-term system performance.
Third, the problem is NP-Hard. Even if the resource allocation $\mathcal{G}$ is determined, the remained task assignment $\mathcal{A}$ is binary.
Considering the optimization space of $\mathcal{A}$ covers all the tasks and vehicles, the optimization problem turns out to be NP-Hard~\cite{hou2020reliable}. 



\section{The Proposed Solution}
\label{sec:alg}
    %
In this section, we propose the \emph{CAVE} algorithm to efficiently address the above optimization problem.
First, we reformulate the original problem into two subproblems at different locations, i.e., task assignment in the ego vehicle and resource allocation in individual passing-by vehicles.
In particular, we tackle the time-correlation issue by incorporating the existing in-progress tasks into both subproblems, where we do not modify their task assignments but re-optimize their resource allocation.
Second, we convert the task assignment subproblem into unconstrained by using the adaptive barrier function, and solve the converted problem by using particle swarm optimization. 
Third, we solve the resource allocation subproblem by deriving the optimal allocation with the Karush–Kuhn–Tucker (KKT) condition.
Note that, the algorithm performs the optimization for incoming tasks while taking care of existing in-progress tasks in both the ego and passing-by vehicles, which addresses the time-correlated issues.

\subsection{Problem Reformulation}
To tackle the coupled optimization variables in the original problem $\mathbb{P}_0$, we decompose it into two subproblems, i.e., task assignment and resource allocation.
The rationale is that, task assignment is performed in the ego vehicle whenever there are incoming tasks from passengers, while the resource allocation is performed in the passing-by vehicles at any time slots.
In particular, to tackle the time-correlation issue, we incorporate existing in-progress tasks into the formulation of the following two subproblems.

On the one hand, when optimizing the task assignment subproblem for incoming tasks, we incorporate the impact from existing in-progress tasks into the objective function.
Denote $\mathcal{K} = \{\mathcal{K}_1, .., \mathcal{K}_j, .., \mathcal{K}_J\}$ as the set of existing in-progress tasks, where $\mathcal{K}_j$ is the subset in the $j$th vehicle. 
Here, we do not re-optimize the task assignment for existing in-progress tasks, as they are already under computation by different passing-by vehicles. 
Thus, we build the task assignment subproblem $\mathbb{P}_1$ in the ego vehicle as 
\begin{align}
    \mathbb{P}_1: \;\; \min _{\mathcal{A}} \;\;\;\;& \sum\limits_{i \in \mathcal{I}+\mathcal{K}} \sum\limits_{j \in \mathcal{J}} \alpha_{ij} L_{i,j} \\
    \text { s.t. } \;\;\;\;& \prod\limits_{j \in \mathcal{J}} (1-\alpha_{i,j}P_j(L_{i,j})) \leq H_i^{min}, \forall i, \label{eq:const_3} 
\end{align}
where we only optimize the task assignment $\mathcal{A}$ for incoming tasks $\mathcal{I}$, although we consider both $\mathcal{I}$ and $\mathcal{K}$.

On the other hand, when optimizing the resource allocation for incoming tasks, we also re-optimize that for existing in-progress tasks.
Thus, we build the resource allocation subproblem $\mathbb{P}_2$ in the passing-by vehicles as  
\begin{align}
    \mathbb{P}_2: \;\;  \min _{\mathcal{G}} \;\;\;\;& \sum\limits_{i \in \mathcal{I}+\mathcal{K} } \sum\limits_{j \in \mathcal{J}} \alpha_{ij} L_{i,j} \\
    \text { s.t. } \;\;\;\;& \sum\limits_{i \in \mathcal{I}+ \mathcal{K}} \alpha_{ij} g_{i,j} \leq G_j^{\max}, \forall j, \label{eq:const_4} 
\end{align}
where the optimization variables $\alpha_{i,j}$ are given, from the perspective of solving this subproblem.

\subsection{Task Assignment Subproblem}
\label{sec:alg:task_assginment}
In this subsection, we aim to solve the task assignment subproblem $\mathbb{P}_1$ by assigning the incoming tasks to passing-by vehicles only.
Here, we identify the difficulties as 
1) unknown parameters (i.e., uplink and downlink wireless data rate) and non-determined resource allocation $\mathcal{G}$.
2) the NP-Hardness of the subproblem with constraints.

\textbf{Parameter Prediction.}
First, we deal with unknown parameters in the subproblem $\mathbb{P}_1$.
Specifically, we create a simple prediction model for each passing-by vehicle to estimate its wireless data rate over time, which will be trained with all historical observations.
The prediction model will observe the number of in-progress tasks and historical wireless data in the passing-by vehicle, and generate the prediction of the next wireless data rate.
In this way, the impact of in-progress tasks will be considered in optimizing incoming tasks.
On the other hand, as resource allocation is performed more frequently after the task assignment, it is difficult to forecast how many computing resources will be allocated to each user task.
To balance the accuracy and complexity, we simply presume the resource allocation will be equally allocated, which will derive the fixed $g_{i,j}$, given the in-progress tasks in individual passing-by vehicles.

\textbf{Adaptive Barrier Method.}
Second, we deal with the constraints in the subproblem $\mathbb{P}_1$, under determined resource allocation and predicted parameters.
Specifically, we use adaptive barrier function inspired by the interior point method~\cite{boyd2004convex} to convert the constrained subproblem into unconstrained.
The basic idea is to adaptively incorporate the constraint into the objective function and then solve a series of unconstrained subproblems.
Hence, we build the unconstrained subproblem as
\begin{align}
    \mathbb{P}_3: \min _{\mathcal{A}} \;\;&\sum\limits_{i \in \mathcal{I} + \mathcal{K}} \sum\limits_{j \in \mathcal{J}} \alpha_{ij} L_{i,j}  \nonumber \\
     & + \mu \sum\limits_{i \in \mathcal{I}} ln \left[H_i^{min} - \prod\limits_{j \in \mathcal{I}} (1-\alpha_{i,j}P_j(L_{i,j})) \right], \label{eq:unconstrained_subproblem} 
\end{align}
where $\mu$ is the non-negative factor, and we use $ln(\cdot)$ penalize if the constraint is violated.
Note that, we consider the impact from existing in-progress tasks in the objective function, while their reliability constraints are skipped.

\textbf{Particle Swarm Optimization.}
Third, we deal with the NP-Hardness in the above unconstrained subproblem $\mathbb{P}_3$ by using particle swarm optimization (PSO)~\cite{eberhart1995new}.
PSO is an efficient global searching algorithm, and has been applied and evaluated in a wide range of application domains~\cite{gad2022particle}, e.g. energy sector and transportation systems. 
Generally, PSO first initializes a candidate solution and iteratively improves the candidate solution (aka. particles) by moving them towards the global optima as well as the local optima among particles in current iteration 
based on a given quality measurement of solutions Eq.~\ref{eq:unconstrained_subproblem}.

Specifically, we first generate the initial task assignment by randomly sampling from its optimization space.
We observe that, to achieve the given requirement of computing reliability, only a partial of vehicles are needed in most scenarios.
Hence, accelerate the convergence of the PSO searching by reducing its optimization space into $n$ vehicles, such as 5 or 10.
Second, as the particle moves during the search iterations, some particles may not satisfy the requirement anymore.
Hence, we dynamically delete some particles and re-sample them again from the reduced optimization space.
Finally, we stop the searching if reaching the given maximum iterations, where the best assignment strategy is chosen according to the quality measurement throughout the whole search iterations.
Note that, PSO can easily be implemented in parallel for further computing acceleration, which would benefit the real-time decision of task assignment.

\subsection{Resource Allocation Subproblem}
In this subsection, we aim to solve the resource allocation subproblem $\mathbb{P}_2$ in individual passing-by vehicles.
Given the task assignment, we observe that the subproblem $\mathbb{P}_2$ is fully separable, with respect to each passing-by vehicle. 
Hence, we apply the KKT condition to derive the optimal resource allocation in each vehicle.
This is based on our observation that, the subproblem is convex by evaluating the Hessian matrix, which are all positive. 

First, we build the Lagrangian function for the $j$th vehicle by including the constraint in Eq.~\ref{eq:const_4} as
\begin{align}
    \mathcal{L}_j = \sum\limits_{i \in \mathcal{I} + \mathcal{K}} \alpha_{ij} L_{i,j}  + \lambda_j (\sum\limits_{i \in \mathcal{I}+ \mathcal{K}} \alpha_{ij} g_{i,j} - G_j^{\max}), \label{eq:lagrangian} 
\end{align}
where $\lambda_j$ is the Lagrange multiplier for the $j$th vehicle.

Then, we differentiate $\mathcal{L}_j$ with respect to $g_{i,j}$ and set the derivative equal to zero, which is expressed as
\begin{align}
    \frac{\partial \mathcal{L}_j}{\partial g_{i,j}} = - \sum\limits_{i \in \mathcal{I}+\mathcal{K}} \frac{\alpha_{i,j} C_i}{ (g_{i,j})^2} + \lambda_j \sum\limits_{i \in \mathcal{I}+ \mathcal{K}} \alpha_{i,j} = 0, \label{eq:derivative}
\end{align}
where the following condition must be satisfied for the inequality constraints, i.e., 
\begin{align}
    \lambda_j (\sum\limits_{i \in \mathcal{I}+ \mathcal{K}} \alpha_{ij} g_{i,j} - G_j^{\max}) = 0. \label{eq:inequality}
\end{align}
Based on Eq.~\ref{eq:derivative} and Eq.~\ref{eq:inequality}, we can obtain that the optimal resource allocation at the $j$th vehicle for both incoming and in-progress tasks, expressed as
\begin{align}
    g_{i,j} = {\sqrt{C_i} G_j^{\max}} /{ \sum\limits_{i \in \mathcal{I}+ \mathcal{K}} \alpha_{ij} \sqrt{C_i}}, \label{eq:opt_resource}
\end{align}
when $\alpha_{i,j}$ is non-zeros, otherwise $g_{i,j} = 0$.


\subsection{The CAVE Algorithm}
Based on the above analysis, we propose the \emph{CAVE} algorithm to solve the problem $\mathcal{P}_0$, whose pseudocode is summarized in Alg.~\ref{alg}.
On the ego vehicle side, it optimizes the task assignment once incoming tasks arrive.
First, we estimate the unknown parameters based on their historical data points.
Second, we initialize the particles and continuously move them in PSO, where particles will be deleted and resampled if the reliability constraint cannot be satisfied.
Third, we stop the PSO searching under given iterations and use the historically best $\mathcal{A}^*$ as the task assignment. 
On the passing-by vehicle side, each of them optimizes its resource allocation at any continuous time (e.g., per milliseconds).
In each passing-by vehicle, we obtain all the tasks and calculate optimal resource allocation based on Eq.~\ref{eq:opt_resource}.

\begin{algorithm}[!t]
    \caption{The \emph{CAVE} algorithm} \label{alg}
    \KwIn{$H_i^{min}, G_j^{\max}, C_i, D_i, E_i, \mu$}
    \KwOut{$\mathcal{A},\mathcal{G}$}

    $/**\;Ego\; Vehicle\; Side **/$\;
    Estimate unknown parameters $R^u_{i,j}, R^d_{i,j}$ and $g_{i,j}$\;
    Initialize $N$ particles by random sampling $\mathcal{A}$\;
    \For{$t = 0,1,...,M$}{
        Calculate Eq.~\ref{eq:unconstrained_subproblem} for all particles\;
        \For{$n = 0,1,...,N$}{
            \If{Eq.~\ref{eq:const_3} is not satisfied}{
                Delete the particle and re-sampling.
            }
        }
        Move particles with PSO accordingly\;
        Decaying $\mu$\;
    }
    Find the best $\mathcal{A}^*$ from historical PSO searching\;

    $/**\;Passing-by\; Vehicle\; Side **/$\;
    Obtain all tasks in each passing-by vehicle\;
    \For{$t = 0,1,2,...$, (parallel)}{
        Calculate optimal resource allocation, $\mathcal{G}_j^* \leftarrow $ Eq.~\ref{eq:opt_resource}\;
    }
    \Return $\mathcal{A}^*$ and $\mathcal{G}_j^*, \forall j$\;
\end{algorithm}

\section{Performance Evaluation}

\begin{figure*}[!t] 
\captionsetup{justification=centering}
  \begin{minipage}[t]{0.33\textwidth}
	\centering
	\includegraphics[width=2.4in]{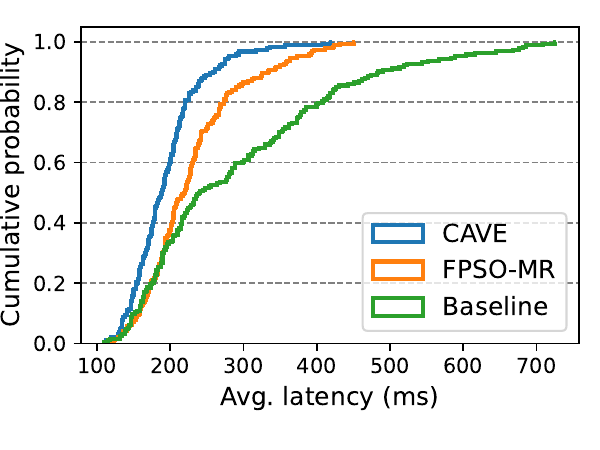}
	\vspace{-0.2in} \caption{\small CDF of task latency under different methods.}
	\label{fig:main_latency}
  \end{minipage}
  \begin{minipage}[t]{0.33\textwidth}
	\centering
	\includegraphics[width=2.4in]{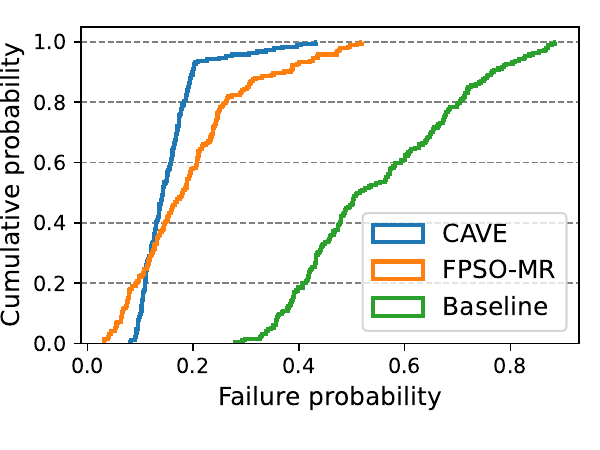}
	\vspace{-0.2in} \caption{\small CDF of task failure probability under different methods.}
	\label{fig:main_failure}
  \end{minipage}
  \begin{minipage}[t]{0.33\textwidth}
	\centering
	\includegraphics[width=2.4in]{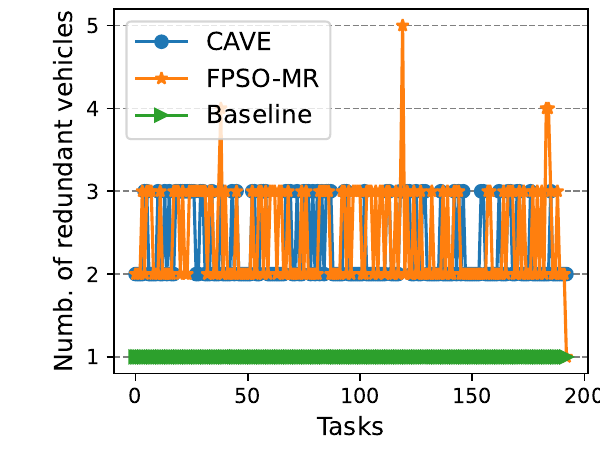}
	\vspace{-0.2in} \caption{\small Number of redundant vehicles per task under different methods.}
	\label{fig:redundant}
  \end{minipage}
\end{figure*}

\textbf{End-to-End Simulator.}
We build a simulator based on our prior work~\cite{xue2023comap} with multiple modules, including ego vehicle computing, downlink wireless network, passing-by vehicle computing, and uplink wireless network.
The basic idea of the time-slotted simulator is to flow \emph{tasks} among these sequential modules, e.g., if a \emph{task} finishes its downlink transmission, it will be enqueued into the next module (i.e., passing-by vehicle computing).
The computation and transmission of \emph{tasks} are simulated by deducting its remaining computing complexity and transmission data size, respectively.
In particular, the wireless network is simulated with an open-source 5G system-level simulator~\cite{oughton2019open}, with the radio channel of urban micro (UMi - Street Canyon).

\textbf{Simulation Parameters.}
The default number of passing-by vehicles is 20, where their locations are randomly generated in a radius of 100 meters.
The default uplink and downlink bandwidth is 10 MHz, with a maximum transmit power of 20 dBm. 
We use the Poisson Point Process (PPP) to generate the user tasks in the ego vehicle, where the default intensity is \textcolor{black}{20} 
per second.
Without loss of generality, transmission size and computation complexity of user tasks are uniformly sampled from [10, 100] Kbits and [1000, 2000] GFLOP.
By default, the computation capacity of vehicles is 10 TFLOPS, and reliability requirement $H^{min}_i=0.2, \forall i$.
In addition, we use the exponential decaying function to represent the reliability function over time, where $P_j(x) = exp(-x), \forall j$, and the unit of $x$ is second.

\textbf{Comparison Algorithms.}
We compare the \emph{CAVE} algorithm with the following works:
\begin{itemize}
    \item \textbf{\emph{Baseline}} uses least-workload criteria to assign user tasks to passing-by vehicles, where computing resources in vehicles are equally shared by all assigned tasks.
    \item \textbf{\emph{FPSO-MR}}~\cite{hou2020reliable} is a fault-tolerant particle swarm optimization algorithm. As its problem is different from ours, we adopt its algorithm idea to solve our problem, where existing in-progress tasks are not considered during its optimization. Besides, computing resources in vehicles are equally shared by all assigned tasks. 
\end{itemize}

\textbf{Latency.}
Fig.~\ref{fig:main_latency} shows the empirical CDF of the average latency of user tasks under different methods.
As we can see that, our proposed \emph{CAVE} algorithm achieves the best latency performance (\textcolor{black}{194.83}ms on average), with a \textcolor{black}{15.25\%} and \textcolor{black}{33.60\%} reduction than the \emph{FPSO-MR} and \emph{Baseline} method, respectively.
Moreover, the \emph{CAVE} algorithm obtains more than 80\% tasks have the latency below \textcolor{black}{221}ms, where the comparative percentile latency is \textcolor{black}{273}ms and \textcolor{black}{408}ms under \emph{FPSO-MR} and \emph{Baseline} method, respectively.
In particular, we observe the long tail of the \emph{Baseline} method, which suggests the necessity of optimizing the task assignment, rather than simply finding the least workload vehicle.

\textbf{Reliability.}
Fig.~\ref{fig:main_failure} shows the empirical CDF of the failure probability of user tasks under different methods.
We can see that, all methods cannot assure 100\% failure probability below the given threshold $H^{min} = 0.2$.
This could be attribute to the complex system dynamics (e.g., traffic, communication, and computing), especially the lack of fully controllability of wireless transmissions.
For example, the \emph{FPSO-MR} method obtains only reliability of 57th percentile for all the tasks (i.e., below 0.2), and the \emph{Baseline} method fails to finish any tasks under the threshold $H^{min} = 0.2$.
In contrast, our proposed \emph{CAVE} algorithm obtains more than 90th percentile for all the tasks (i.e., below 0.2), with a maximum of \textcolor{black}{0.431} failure probability. 

\textbf{Redundancy.}
To dissect the details behind the failure probability, we show the number of redundant assignment of user tasks over time in Fig.~\ref{fig:redundant}, under different methods.
The \emph{Baseline} method always select only one vehicle, and the \emph{FPSO-MR} method assign a task to \textcolor{black}{2.45} 
vehicles on average.
In contrast, the \emph{CAVE} algorithm tends to assign more vehicles (average \textcolor{black}{2.33} 
vehicles), to assure the given threshold of the failure probability.
\textcolor{black}{We observe that, although the \emph{FPSO-MR} method assigns tasks to more vehicles, its achieved latency performance is still worse than our proposed \emph{CAVE} algorithm.}
This may be attribute to the reason that, it overlooks the in-progress tasks when optimizing the incoming user tasks, which can lead to non-trivial performance degradation of user tasks, including both latency and reliability.



\begin{figure*}[!t] 
\captionsetup{justification=centering}
  \begin{minipage}[t]{0.33\textwidth}
	\centering
	\includegraphics[width=2.4in]{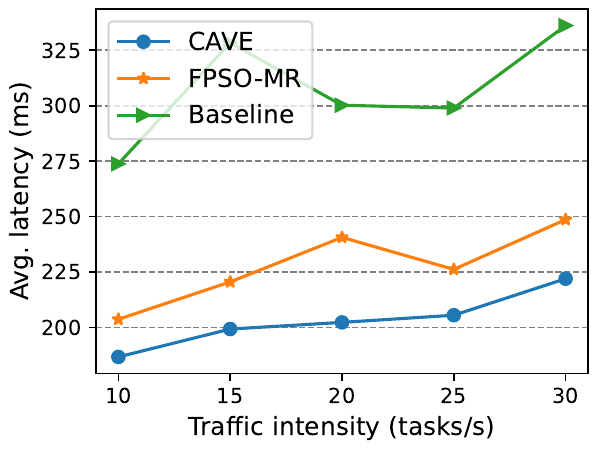}
	\vspace{-0.2in} \caption{\small Average task latency under various traffic.}
	\label{fig:intensity_latency}
  \end{minipage}
  \begin{minipage}[t]{0.33\textwidth}
	\centering
	\includegraphics[width=2.4in]{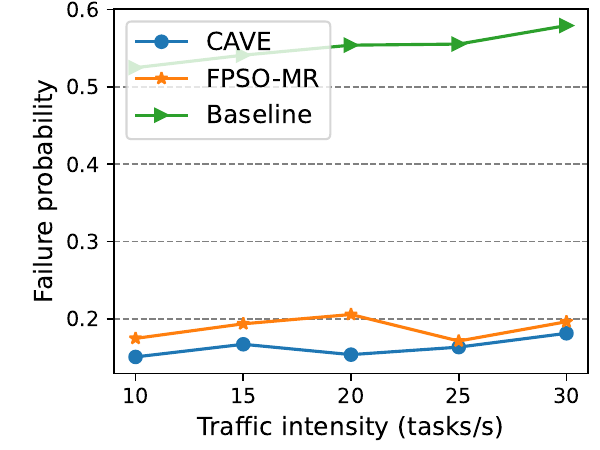}
	\vspace{-0.2in} \caption{\small Task failure probability under various traffic.}
	\label{fig:intensity_failure}
  \end{minipage}
  \begin{minipage}[t]{0.33\textwidth}
	\centering
	\includegraphics[width=2.4in]{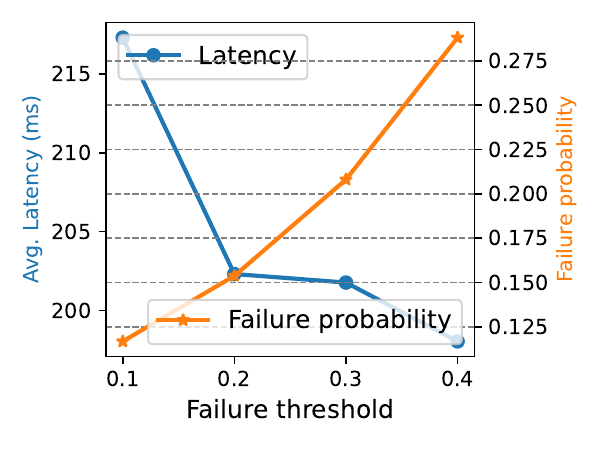}
	\vspace{-0.2in} \caption{\small \emph{CAVE} performance under different failure thresholds.}
	\label{fig:threshold}
  \end{minipage}
\end{figure*}

\textbf{Traffic Intensity.}
Fig.~\ref{fig:intensity_latency} and Fig.~\ref{fig:intensity_failure} show the average latency and failure probability of user tasks under various traffic intensities, respectively.
As more user tasks in the system, the average task latency of all methods generally increases, because the resource competition becomes more severe in both communication and computation.
In the \emph{CAVE} algorithm, the average latency is increased from \textcolor{black}{186.1}ms to \textcolor{black}{221.4}ms, and the failure probability also increased, but still not exceed the given threshold of 0.2.
In contrast, we observe the other methods are with much higher task latency. 
This result justifies the \emph{CAVE} algorithm in handling varying user traffic and assuring the computing reliability. 


\textbf{Failure Threshold.}
Fig.~\ref{fig:threshold} shows the average latency and failure probability of user tasks under different failure thresholds, respectively.
The lower failure threshold means that the more redundant vehicles per task needed to assure the computing reliability.
Hence, we found that the average number of redundant vehicles per task increase from
\textcolor{black}{1.62 ($H^{min}=0.4$) to 2.96 ($H^{min}=0.1$)} 
in the \emph{CAVE} algorithm.
As a result, the latency performance generally increases under lower failure thresholds.
This results show that the \emph{CAVE} algorithm can adapt to different reliability functions while assuring the reliable computing for user tasks.

\section{Related Work}
\label{sec:prior}
This work relates to the computation offloading, resource allocation, and reliable computing in the scenario of vehicle edge computing.
Computation offloading~\cite{feng2022latency, xue2023comap} is the widely used approach to exploit more powerful servers (e.g., edge or cloud) to accelerate the task computation of mobile devices and vehicles.
To achieve diverse objective (e.g., latency and energy), existing works have formulated various optimization problems and designed a wide range of algorithms and methods, via both model-based and model-free approaches.
Task assignment and resource allocation are extensively investigated and optimized to achieve more efficient and effective computation offloading~\cite{chen2021efficient}.
For example, Feng \emph{et. al.}~\cite{feng2022latency} proposed two algorithms that minimize the system latency under both binary and partial reverse offloading problems respectively, by optimizing both offloading decision and radio resource allocation.
However, most existing works~\cite{xue2023comap} focused on offloading tasks from vehicles to infrastructural edge/cloud servers, where the wireless connectivity is more reliable and consistent.
In the scenario of in-vehicle edge computing, we explore the idea of crowdsourcing passing-by vehicles to serve passengers tasks, where the vehicle-to-vehicle wireless connectivity becomes much more volatile over time.
Several works~\cite{hou2020reliable,liu2023accelerating} focused on the reliable computing in vehicle edge computing, by assigning single task to multiple vehicles to increase computation redundancy.
However, their works considered only snapshot-based task computation, where the complex time correlation among consecutive tasks in individual vehicles are overlooked.
In this paper, we focus on the reliable computing problem in the in-vehicle edge computing, where the time correlation of tasks are incorporated in the design of the \emph{CAVE} algorithm.



\section{Conclusion}
In this work, we explored the possibility of crowdsourcing passing-by vehicles to accelerate the computation of the ego vehicle.
We formulate the optimization problem to reduce the average end-to-end latency by optimizing task assignment and resource allocation for user tasks.
We designed the \emph{CAVE} algorithm with several key designs to efficiently solve the problem.
Extensive simulation results show that our proposed algorithm outperformed existing solutions, in terms of convergence, reliability, and scalability. 


\bibliographystyle{IEEEtran}
\bibliography{ref/qiang, ref/ref}

\end{document}